\documentclass[aps,twocolumn,noshowpacs,superscriptaddress]{revtex4}
\usepackage{amsmath,graphicx,amsbsy,amssymb,amsmath,epsfig,latexsym,wasysym,pifont,mathrsfs,natbib,epstopdf}
\usepackage{float}
\newfloat{widefig}{thp}{lop}
\usepackage{comment}
\usepackage{verbatim}
\usepackage{multirow,array}
\usepackage{hyperref}
\usepackage{color}

\begin{document}
\newcommand{\mycomment}[1]{\textbf{\color{green}{#1}}}

\title{Interplay of phase sequence and electronic structure in the modulated martensites of Mn$_2$NiGa from first-principles calculations}


\author{Ashis Kundu}
\email[]{k.ashis@iitg.ernet.in}
\affiliation{Department of Physics, Indian Institute of Technology Guwahati, Guwahati-781039, Assam, India.}
\author{Markus E. Gruner}
\email[]{Markus.Gruner@uni-due.de}
\affiliation{Faculty of Physics and Center for Nanointegration, CENIDE,
  University of Duisburg-Essen, D-47048 Duisburg, Germany}
\author{Mario Siewert}
\affiliation{Faculty of Physics and Center for Nanointegration, CENIDE,
  University of Duisburg-Essen, D-47048 Duisburg, Germany}
\author{Alfred Hucht}
\affiliation{Faculty of Physics and Center for Nanointegration, CENIDE,
  University of Duisburg-Essen, D-47048 Duisburg, Germany}
\author{Peter Entel}
\email[]{Peter.Entel@uni-due.de}
\affiliation{Faculty of Physics and Center for Nanointegration, CENIDE,
  University of Duisburg-Essen, D-47048 Duisburg, Germany}
\author{Subhradip Ghosh}
\email[Corresponding author:]{subhra@iitg.ernet.in}
\affiliation{Department of Physics, Indian Institute of Technology Guwahati, Guwahati-781039, Assam, India.}

\date{\today}
 
\begin{abstract}
  We investigate the relative stability, structural properties and electronic structure
  of various modulated martensites of the magnetic shape memory alloy Mn$_{2}$NiGa by means of density functional theory.
  We observe that the instability in the high-temperature cubic structure first drives the system
  to a structure where modulation shuffles with a period of six atomic planes are taken into account.
  The driving mechanism for this instability is found to be the nesting of the minority band Fermi surface,
  in a similar way to that established for the prototype system Ni$_{2}$MnGa.
  In agreement with experiments, we find 14M modulated structures with orthorhombic and monoclinic symmetries
  having energies lower than other modulated phases with the same symmetry.
  In addition, we also find energetically favourable 10M modulated structures which have not been observed
  experimentally for this system yet.
  The relative stability of various martensites is explained in terms of changes in the electronic structures
  near the Fermi level, affected mostly by the hybridisation of Ni and Mn states.
  Our results indicate that the maximum achievable magnetic field-induced strain in Mn$_{2}$NiGa would be
  larger than in Ni$_{2}$MnGa. However, the energy costs for creating nanoscale adaptive twin boundaries are found to be one
  order of magnitude higher than that in Ni$_{2}$MnGa.
\end{abstract}

\maketitle

\def\vvec#1{\underline{#1}}

\section{INTRODUCTION}
Magnetic shape memory (MSM) alloys have drawn much attention in recent years  due to substantial magnetic
field induced strain (MFIS) and large magnetocaloric effect (MCE), which are useful for
state of the art
technologies~\cite{UllakkoAPL96,MurrayAPL00,SozinovAPL02,ChmielusNM09,MarcosPRB02,HuPRB01,KrenkePRB07,PasqualePRB05}.
The magneto-structural coupling, along with magnetocrystaline anisotropy, increases the MFIS,
thus making the material interesting for magnetomechanical actuators~\cite{MarcosPRB02}.
The observation of a large reversible MFIS is connected to the presence of modulated low-symmetry martensites
which evolve when the high-temperature, high-symmetry cubic structure phase
transforms to low-temperature, low-symmetry structures in a diffusionless way~\cite{ChernenkoPRB98}.
These modulated martensites can be
commensurate and incommensurate orthorhombic or twinned monoclinic structures~\cite{BrownJPCM02,Pons05,RighiAM07,RighiAM08,SinghPRB14,MariagerAM14}. The phases exhibit a high mobility of particular twin boundaries~\cite{StrakaAM11,DiestelAPL11},
which can even be driven in moderate magnetic fields and a
low hysteresis which makes them appealing as smart materials for actuator and sensor applications.
This is believed to be a consequence of the formation of a hierarchical martensitic microstructure, which involves
twinning of twinned or modulated structures, spanning several length scales~\cite{MuellnerJMMM03,KaufmannPRL10,MuellnerAM10}.
Depending on the martensitic structure at the working temperature, different manifestations of the functional
properties can be achieved in these systems. For example, a 6$\%$ strain was obtained in the martensitic phase
of Ni-Mn-Ga alloy with a five-layered modulated (10M) structure~\cite{MurrayAPL00,SozinovIEEE02}.
Later MFIS as large as 10$\%$ was observed in the same system  with a seven-layered modulated
structure (14M)~\cite{SozinovAPL02,SozinovIEEE02}.

A volume of work has been done on the prototype Ni-excess Ni-Mn-Ga system, often in the stoichiometric composition of 2:1:1,
to understand the origin of these modulated structures.
One line of thought considers solely the softening of a particular vibrational mode at a specific wave vector,
due to Fermi surface nesting, as the origin of the modulated phases, which are shearing
of lattice planes within the equilibrium martensite phase~\cite{LeePRB02,BungaroPRB03}.
In contrast, the concept of adaptive martensites~\cite{adaptivePRB91}
considers the modulated structures, instead of being equilibrium structures,
as metastable microstructures of the nonmodulated (NM) martensite.
Both concepts gained significant currency due to corroborations by theoretical
calculations and experimental observations, e.g., for the case of stoichiometric and Ni-excess Ni-Mn-Z systems~\cite{PonsActamater2000,KaufmannPRL10,KaufmannNJP11,NiemannAEM12,DuttaPRL16,MartinPRB16}. Very recently, an attempt  has been made to unify the opposing lines of
thought~\cite{Gruner2017}.
Nevertheless, the fundamental understanding of the modulated phases can only be considered complete
if the underlying concepts can be applied to a broader set of systems.

Although Ni$_{2}$MnGa, the prototype MSM, has wonderful prospects for several applications, it has some serious limitations.
The martensitic transformation temperature (T$_m{}$) of stoichiometric Ni$_2{}$MnGa is only around 200 K, while
the Curie temperature (T$_c{}$) is nearly 380 K, which is still too close to room temperature~\cite{WebsterPMB84,OpeilPRL08}.
These properties can be improved in off-stoichiometric systems but
the use as a magnetic field induced actuator is -- apart from the low value of T$_m{}$ -- also restricted due to brittleness,
high dependence of MFIS on the crystal structure.
Attempts to discover other MSMs which would take care of these  shortcomings, led to Mn$_2{}$NiGa~\cite{LiuAPL05}.
Mn$_2{}$NiGa shows martensitic transformation near room temperature (T$_m{}$=270 K) and has a high
Curie temperature (T$_c{}$=588 K) which make this material promising for better practical applications~\cite{LiuAPL05,LiuPRB06}.
The ferrimagnetic coupling between two Mn atoms gives rise to further interesting physical properties
in Mn$_2{}$NiGa~\cite{BarmanPRB08}.
Singh {\em et al.} reported spin-valve-like magnetoresistance in Mn$_2{}$NiGa at room temperature~\cite{SinghPRL13}.
Recently, a notable inverse magnetocaloric effect (MCE) was reported in Mn$_2{}$NiGa~\cite{SinghAPL14}.
Liu {\em et al.} found that Mn$_2{}$NiGa exhibits 4$\%$ MFIS while its low-temperature  structure is
nonmodulated tetragonal~\cite{LiuAPL05}.
Indeed, total energy calculations on tetragonally deformed Heusler structures by density functional theory (DFT)
based methods showed that Mn$_2{}$Ni-based Heusler alloys undergo volume-conserving transformations
from cubic to tetragonal structures~\cite{PaulJAP11,WollmannPRB15}.
However, modulated structures are expected when a magnetic system undergoes diffusion-less martensitic transformation
and also exhibits MFIS, and a seven-layered monoclinic modulated structure was observed at room
temperature in the  powder x-ray diffraction study of Mn$_2{}$NiGa~\cite{SinghAPL10}.
It was observed that the structure of the martensitic variant is quite sensitive to the residual stress in the system.
Recent neutron diffraction study of Mn$_2{}$NiGa has also reported a seven-layered orthorhombic modulated structure
at low temperature~\cite{BrownJPCM10}.
The handful of experimental results and the signatures obtained from the DFT calculations thus call for a detailed investigation into the  energy landscape to understand and interpret the  origin and stabilities of different modulated structures in the martensite phase of Mn$_2$NiGa. Such an investigation is required since the efficiency of MFIS and MCE  depend substantially on the structures in the martensite phases for MSM systems. Such a study would also provide further insights into the mechanism of the martensitic phase transformation in Ni-Mn-Ga systems and put the research on understanding of the modulated structures in Heusler  MSM compounds in a broader perspective.

In this paper, we therefore investigate the structural properties and relative stabilities of different modulated structures of Mn$_2{}$NiGa and make an attempt to understand the origin of the sequences of structural phases as the system is driven from high-temperature cubic to low-temperature NM tetragonal variant. Comparisons with Ni$_{2}$MnGa are made in order to understand the physical mechanism. Using DFT calculations, we first perform an elaborate study of the structural properties of the various crystalline phases. Then we present and discuss results on the sequences of various phases and their relative stabilities from their electronic structures. The origin of the modulated structures is explained next by presenting results on phonon dispersions, the features in the Fermi surfaces, and the electronic susceptibilities.
\begin{figure*}[t]
\centerline{\hfill
\psfig{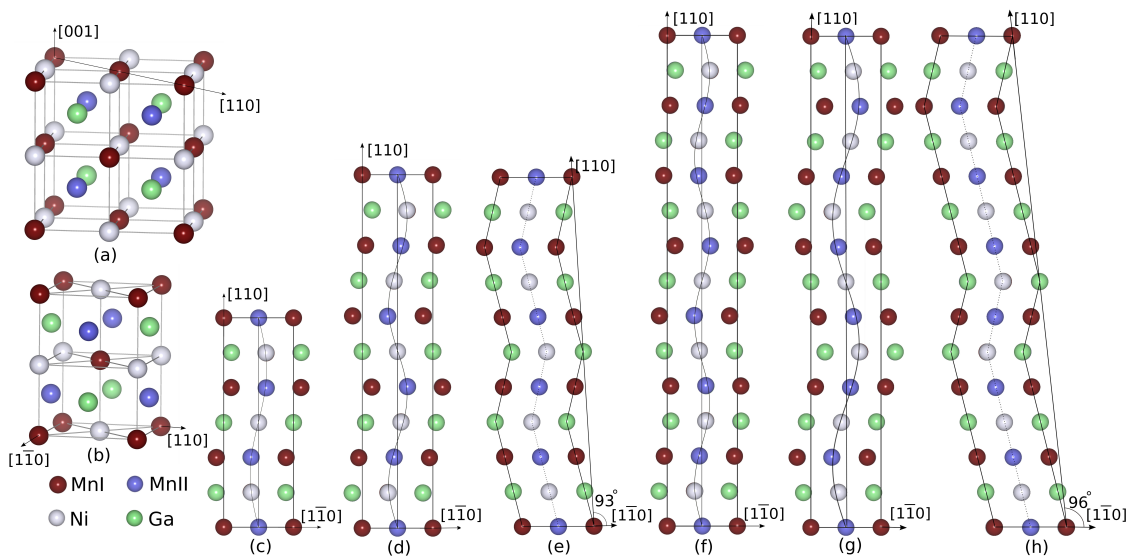}\hfill}
\caption{(a) Cubic Hg$_{2}$CuTi structure of Mn$_2{}$NiGa. (b) Non modulated tetragonal (L1$_{0}$)structure. (c) 6M structure. (d) 10M structure. (e) 10M$(3\bar{2})_2$  structure. (f) 14M$_{3/7}$ structure. (g) 14M$_{2/7}$ structure. (h) 14M$(5\bar{2})_2$  structure for the martensitic phase.}
\label{struct}
\end{figure*}
\section{COMPUTATIONAL DETAILS}
Electronic structure calculations were performed with the spin-polarized density functional theory (DFT) based  projector augmented wave method~\cite{PAW94} as implemented in the Vienna \textit{ab initio} Simulation Package (VASP)~\cite{VASP196,VASP299}. For all calculations, we have used the PBE-GGA functional~\cite{PBEGGA96} for exchange and correlation parts in the Hamiltonian. A high kinetic energy cutoff of 750 eV was used for all calculations to address the small energy difference between different modulated structural phases.
For Ga, the 3$d$ electrons were included in the valence band. The Brillouin zone was sampled by the Monkhorst-Pack~\cite{MP89} $k$-point generation scheme with a uniform  21x21x21 $k$-point mesh for the cubic structure, a 12x4x8 mesh for 6M modulated structures, and a 10x2x8 mesh for 10M and 14M modulated structures, after careful convergence tests. The total energy convergence criteria were  set to $10^{-5}$ eV and the force convergence criteria were set to $10^{-2}$ eV/\r{A}.

Phonon dispersion relations were calculated with the direct method as implemented in the PHON package~\cite{DarioCPC09}. A 4x4x4 supercell containing 256 atoms with a small atomic displacement of 0.02 \r{A} was used to calculate the force constant matrices.
An energy cut-off of 500 eV and a 2x2x2 $k$ mesh  were used  for the calculations of forces.

Fermi surfaces and generalized susceptibilities were obtained from electronic structure calculations. As for the total energy calculations, we employed the VASP code using the PBE functional and the tetrahedron method for the integration over a dense $\Gamma$-centered $k$-mesh containing $75\times 75\times 75$ points in the full Brillouin zone (9880 in the irreducible part). This dense mesh was used in the self-consistency cycle to determine charge distribution and eigenvalues which are required to be significantly more accurate than in our previous calculations of the Fermi surfaces of Ni$_2$MnGa obtained by a similar procedure~\cite{EntelMSF08,Siewert12AEM}. The $k$-dependent eigenvalues were processed by a script written for the computer algebra system Mathematica. The script unfolds the irreducible part of the Brillouin zone and determines the Fermi surface as the cross section at the Fermi level of the four-dimensional energy in reciprocal space, interpolated by cubic splines. The real part of the generalized susceptibility was obtained from the same quantity in the way described in Sec. IIID.

\section{Results and Discussions}
\subsection{Structural properties of cubic, non-modulated, and different modulated phases}

In this subsection, we present detailed results on the structural parameters of the different phases
of Mn$_{2}$NiGa at different temperatures. The high-temperature phase of this system, like
Ni$_{2}$MnGa, is a variant of the L2$_{1}$ structure.
Experimental results~\cite{LiuAPL05,LiuPRB06,SinghAPL10} and  theoretical calculations~\cite{BarmanPRB08,PaulJAP11,WollmannPRB15} have confirmed that the the high-temperature, high symmetry structure of Mn$_2{}$NiGa alloy is the Hg$_2{}$CuTi structure (space group No. 216; $F\bar{4}3m$) with four inequivalent Wyckoff positions (4a, 4b, 4c, 4d) in the unit cell. The 4a (0,0,0) and 4c (0.25, 0.25, 0.25) Wyckoff positions are occupied, respectively, by two Mn atoms, MnI and MnII.
Ni and Ga occupy the 4b (0.5, 0.5, 0.5) and 4d (0.75, 0.75, 0.75) sites respectively.
The structure is shown in Fig.\ \ref{struct}(a). The equilibrium lattice constant is determined by calculating
the total energy as a function of lattice parameters and then  by fitting the results to the Murnaghan equation of state.
Our calculated lattice constant a=5.84 \r{A}  is within 1$\%$ with the experimental lattice constant a=5.907 \r{A}~\cite{LiuAPL05} and shows excellent agreement with other theoretical results~\cite{BarmanPRB08,PaulJAP11,WollmannPRB15}.


Liu {\em et al.} first reported the shape memory behavior in Mn$_2{}$NiGa and observed a martensitic
transformation around room temperature~\cite{LiuAPL05,LiuPRB06}.
From the x-ray diffraction pattern in the system, they concluded that the low-temperature structure is a nonmodulated(NM) tetragonal structure, as shown in Fig.\ \ref{struct}(b). In order to verify this, we  tetragonally  distort the cubic structure, keeping the volume constant at the equilibrium volume of the high- temperature structure,  and compute the total energy as a function of the tetragonal distortion provided by the value of $c/a$. The total energy curve of Mn$_{2}$NiGa consists of two minima, one shallow at $c/a = 0.93$ and another deep at $c/a = 1.28$, as shown in Fig.\ \ref{en-tot}. The energy difference between the cubic($c/a=1.0$) and NM tetragonal structure at $c/a = 0.93$ is about 3 meV/atom whereas that value for $c/a = 1.28$ is about 28 meV/atom. Our calculated values of the $c/a$  and the energy differences are in good agreement with a previously reported result~\cite{PaulJAP11}.
This, thus is in agreement with the experimental observations from the x-ray diffraction data.
However, the presence of a meta-stable minimum for $c/a<1$ indicates that other martensitic NM structures may be
present in Mn$_{2}$NiGa. A similar feature has also been reported for the related isoelectronic compound Mn$_2$PtGa~\cite{Roy17JMMM}.
For Ni$_2{}$MnGa, a
stable phase with $c/a < 1$ could only be obtained in DFT calculations by shuffling the atomic planes
resulting in modulated structures, such as 6M, 10M, and 14M, which differ from one another
by the period of modulations~\cite{ZayakJPCM03,ZayakPT05,RighiJSSC06,RighiAM07,RighiAM08,SinghPRB14,LuoAM11}.
It has also been observed that  the existence of a particular modulated structure in Ni-Mn-Ga
system depends on the relative concentrations of the three constituents and also on the
resolution of the diffraction technique~\cite{SinghPRB14}.

The results above and the observation of a 14M monoclinic modulated structure by Singh {\it et al.}~\cite{SinghAPL10}
and a 14M orthorhombic modulated structure by Brown {\it et al.} at low temperature~\cite{BrownJPCM10}
in Mn$_{2}$NiGa show that there can be several modulated structures in this system, depending
on the periodicity of the modulations and/or the stacking sequences of atomic layers.
Accordingly we perform detailed computations of the structural aspects of different modulated structures of
Mn$_{2}$NiGa, with orthorhombic and monoclinic lattice parameters. In what follows, we present results on pseudotetragonal 6M, 10M, and 14M
modulated structures with orthorhombic lattice parameters originating from wavelike modulations, and on 10M and 14M
modulated structures with monoclinic symmetry originating from stacking sequences.

The 10M structure in Ni$_{2}$MnGa is probably the most studied
modulated martensite in terms of atomistic simulations
(see, e.g., Refs.\ \onlinecite{ZayakJPCM03,Zayak04,ZayakPRB05,ZayakPT05,Entel06Review,Hickel08,KartPSS08,LuoAM11,HickelAEM12,Gruner2017}). Experimentally, the 10M of Ni$_2$MnGa is frequently reported as an incommensurate
modulation~\cite{RighiJSSC06,RighiAM07,SinghPRB14,Cakir13,MariagerAM14},
e.\,g, its period does not amount to exactly 10 lattice planes, but there is some notable
deviation leading to irrational wave vectors, which is difficult to  model accurately,
as it requires large supercells~\cite{Xu12,Bai13}.
An incommensurate modulation is straightforwardly explained
in terms of a soft phonon mode arising from an electronic instability at the Fermi surface
(which can lead to arbitrary wave vectors), instead of a twinned microstructure.
But there is an ongoing discussion, as to whether a statistically perturbed stacking sequence
might alternatively explain the experimental observation~\cite{MariagerAM14,GrunerPSSB14,SinghPRB2015,Gruner2017}.
Since the differences in total energy are expected to be rather small, it is very difficult to
accurately predict incommensurate modulations from atomistic calculations.
Therefore, we will concentrate, as most theoretical approaches up to now, on the
paradigmatic commensurate modulations ubiquitous in literature, like 6M, 10M and 14M.
These will be considered alternatively as sinusoidal modulations confined in a orthorhombic cell and as nano-twinned structures with fully optimized monoclinic lattice parameters.

Although no report on the existence of a 10M structure is available for Mn$_{2}$NiGa, we performed structural optimization and total energy calculation for this structure.
Following Zayak {\em et al.}~\cite{ZayakJPCM03}, we constructed a $1 \times 5 \times 1$ supercell of a body- centered tetragonal (bct) structure, generated from the high-temperature cubic structure. The supercell, shown in Fig.\ 1(d), contains 40 atoms. The modulation was incorporated by shuffling the $(110)$ atomic planes  along $[1\bar{1}0]$ direction, thus including two sine periods
in the ten atomic layers along $[110]$. An initial displacement of 0.2 \r{A} was chosen for both Ni-Mn and
Mn-Ga atomic planes.

\begin{figure}[t]
\centerline{\hfill
\psfig{file=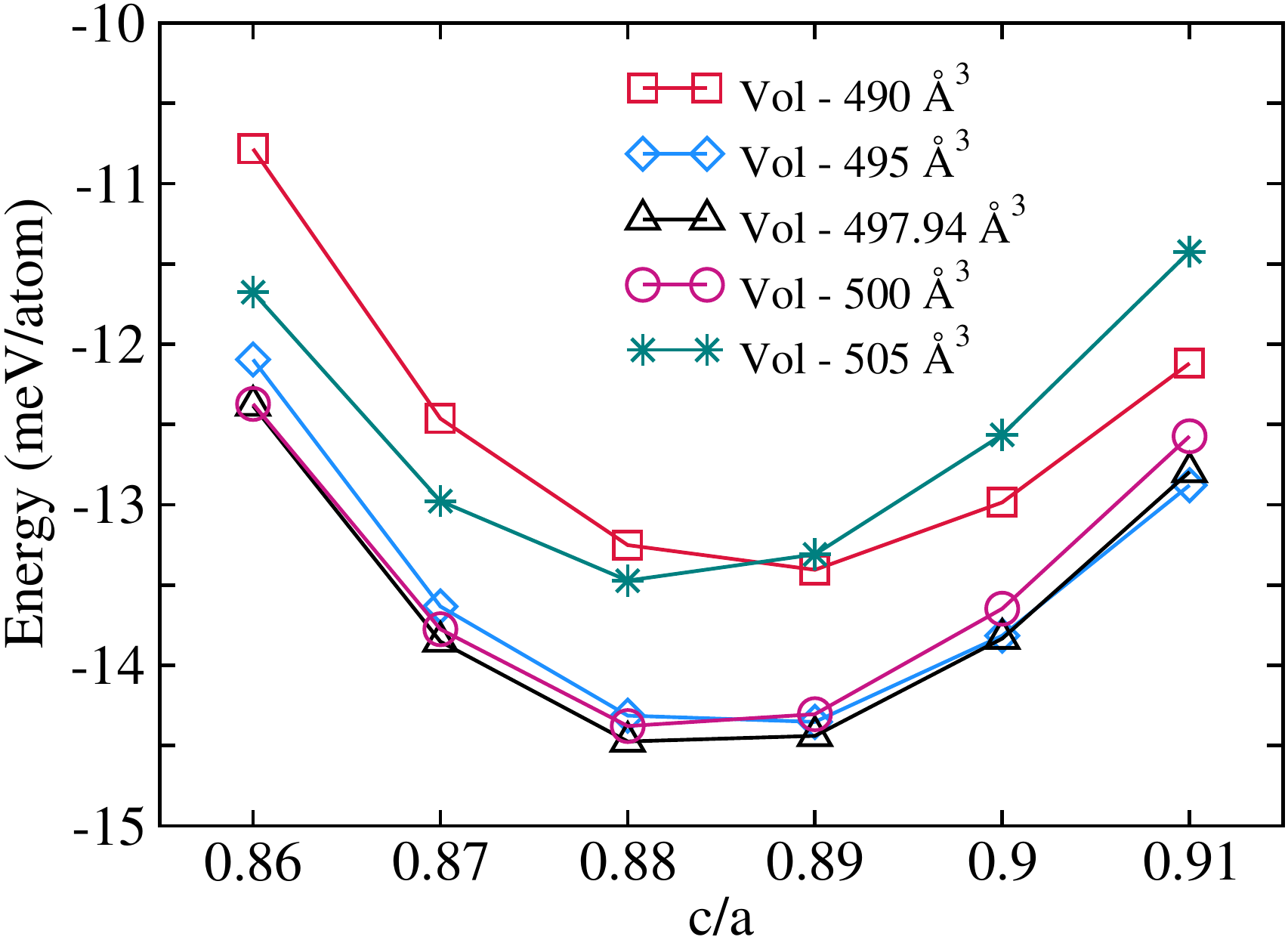,width=0.40\textwidth}\hfill}
\caption{The total energies of pseudotetragonal 10M structure (with orthorhombic lattice parameters)
  relative to the cubic Hg$_{2}$CuTi structure, as a function of $c/a$ ratio for five different volumes:
  490 \AA$^3$, 495 \AA$^3$, 497.94 \AA$^3$(equilibrium volume of cubic structure), 500 \AA$^3$, and 505 \AA$^3$.}
\label{en-vol}
\end{figure}

\begin{figure}[t]
\centerline{\hfill
\psfig{file=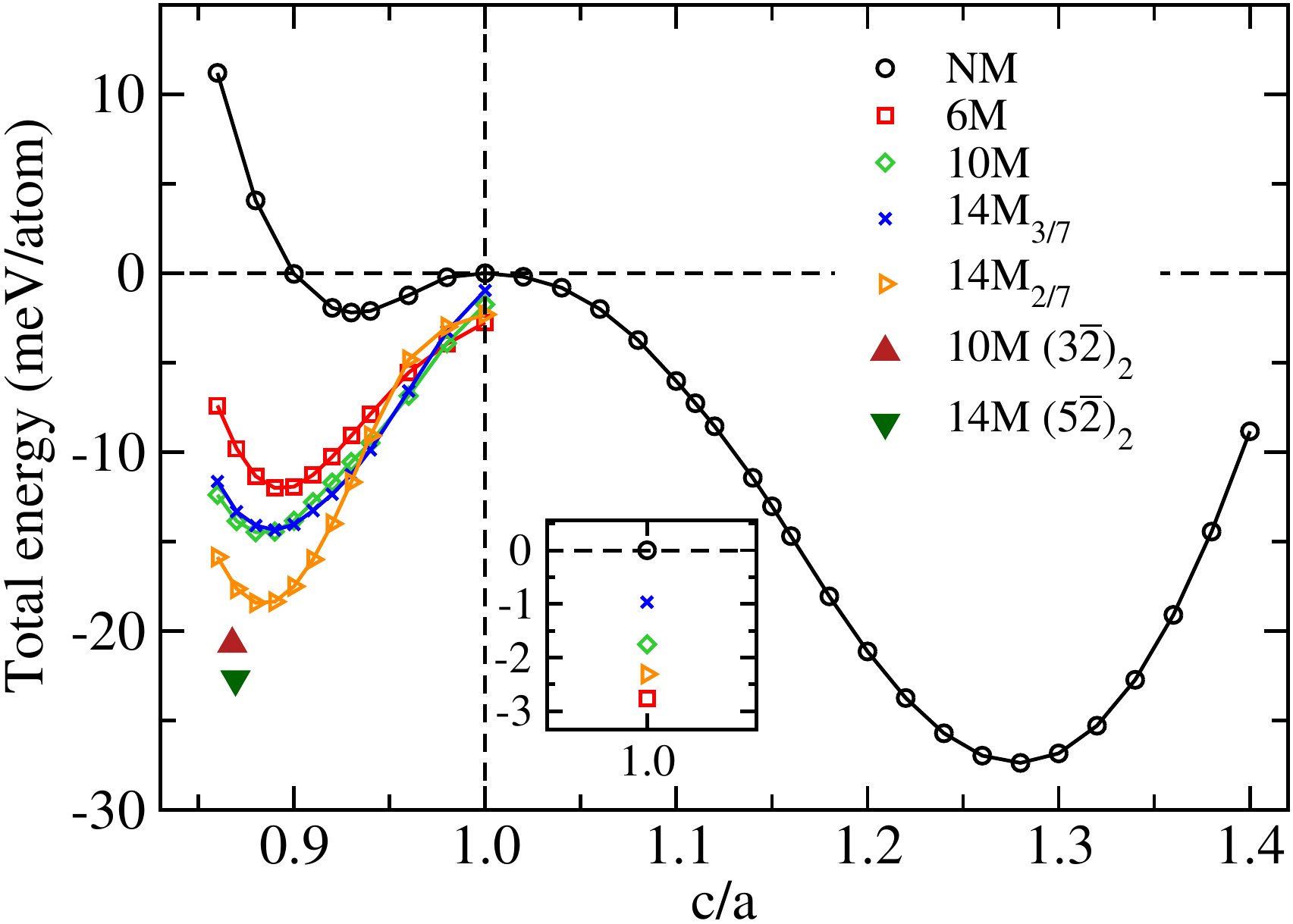,width=0.40\textwidth}\hfill}
\caption{The variation of the total energy  in NM tetragonal (L1$_{0}$) and in the different modulated structures
  of Mn$_{2}$NiGa as a function of the $c/a$ ratio. The zero energy is taken to be the energy of the cubic
  Hg$_{2}$CuTi-type inverse Heusler structure.
  The energies of the pseudo-cubic structures (modulated structures inscribed in an orthorhombic cell
  with $c/a=1$) are shown in the inset.}
\label{en-tot}
\end{figure}

We optimized the atomic positions of 10M structure for different volumes around the equilibrium volume of the high-temperature phase, by keeping the shape of the cell fixed, thus representing the pseudotetragonal or orthorhombic structure. The results of total energies for different volume of the 10M structure are shown in Fig.\ \ref{en-vol}. We obtained an energy minimum at $c/a=0.88$.
From Fig.\ \ref{en-vol}, it is observed that the $c/a$ ratio does not change much with the volume of the cell. Therefore, in order to investigate the volume change, we have calculated the total energies at different volumes, but with the $c/a$ ratio at a constant value of 0.88. We obtained  a volume change of 0.05\% for the 10M structure, (Table\ \ref{table1}). After atomic relaxation the structure preserves the wavelike modulation with modulation amplitude 0.565 \r{A} for the Mn-Ni plane and 0.554 \r{A} for the Mn-Ga plane.

In order to benchmark our technical settings, we also calculated  the structural parameters of the 10M Ni$_2{}$MnGa
structure using the same approach.
We obtained an energy minimum at $c/a= 0.92$ with a modulation amplitude of 0.332 \r{A} for the
Mn-Ga planes and 0.363 \r{A} for the Ni planes.
The obtained c/a ratio of the 10M structure for Ni$_2{}$MnGa is consistent with other theoretical
results~\cite{ZayakJPCM03,KartPSS08,LuoAM11}.
The optimized modulation amplitudes in our calculation are slightly larger than those reported by Zayak {\em et al.}.
This could be due to the $c/a$ ratio of 0.92 obtained in our calculations which is smaller than the value 0.955
used by Zayak {\em et al.}~\cite{ZayakJPCM03}
according to the linear relationship of modulation amplitude with the c/a ratio~\cite{LuoAM11}.
%

\begin{table}[b]
\centering
\caption{\label{table1} The calculated lattice parameters of Mn$_{2}$NiGa in cubic Hg$_{2}$CuTi-type,
  NM tetragonal(L1$_{0}$), 6M, 10M and 14M structures with wavelike modulation. For the pseudo-tetragonal 6M, 10M and 14M, the orthorhombic lattice parameter $b$ is fixed as $3\times a$, $5\times a$, and $7\times a$, respectively, while the ratio
  $c/a_0$ is calculated from $a_0=\sqrt{2}\,a$ allowing direct comparison with the cubic case where $c/a=1$.}
\begin{tabular}{c@{\hspace{0.3cm}}c@{\hspace{0.3cm}}c@{\hspace{0.3cm}}c@{\hspace{0.3cm}}c@{\hspace{0.3cm}}}
\hline\hline \multicolumn{5}{c}{Lattice parameter } \\ \hline
Structure  & a (\r{A}) & c (\r{A}) & c/a$_0$  & $\Delta{V}/V$(\%)   \\ \hline
Cubic       &  5.84(5.90)  & 5.84   & 1.00  &  0.00   \\
6M          &  4.29  &  5.40  &  0.89  &  0.22   \\
10M         &  4.30  &  5.36  &  0.88  &  0.05  \\
14M$_{3/7}$    &  4.29  &  5.40  &  0.89  &  0.22  \\
14M$_{2/7}$    &  4.31  &  5.36  &  0.88  &  0.05  \\
Tetragonal  &  3.80  &  6.88  &  1.28  &  0.65   \\
14M[expt.]~\cite{BrownJPCM10}  &  4.172  &  5.336  &  0.90  &    \\

\hline\hline
\end{tabular}

\end{table}

Next, we  investigated the 6M and 14M structures with orthorhombic symmetry for Mn$_{2}$NiGa. The 6M and 14M structures
were constructed by 1 x 3 x 1 and 1 x 7 x 1 supercells of the bct structure as shown in Figs. \ref{struct}(c), (f) and (g), respectively.
The 6M and 14M structures contain 24 and 56 atoms, respectively, in the unit cell. For the 6M structure,
one period was fitted into six atomic planes.
For the 14M structure, two possible kinds of modulations are reported in the context of Ni$_2{}$MnGa:
(1) two periods fitted into 14 atomic planes, referred to as 14M$_{2/7}$[see Fig. \ref{struct}(f)]~\cite{RighiAM08} and
(2) three periods fitted into 14 atomic planes, referred to as 14M$_{3/7}$[see Fig. \ref{struct}(g)]~\cite{RighiJSSC06}.
In the absence of any experimental information on the nature of the 14M modulation in this system,
we considered both types of 14M modulations in our work.
The optimized lattice parameters for all the structures considered are given in Table \ref{table1}.
The modulation amplitudes of the 6M, 14M$_{3/7}$, and 14M$_{2/7}$ structures are 0.506 \r{A}, 0.5 \r{A} and 0.52 \r{A}, respectively,
for Mn-Ni planes and 0.492 \r{A}, 0.481 \r{A} and 0.507 \r{A}, respectively, for Mn-Ga planes.
Our calculated structural parameters of different modulated structures and the NM tetragonal structure
are within 2-3 $\%$ of the available experimental lattice parameters~\cite{LiuAPL05,BrownJPCM10}.

\begin{table}[b]
\centering
\caption{\label{table2}The calculated lattice parameters of Mn$_{2}$NiGa in 10M and 14M monoclinic structures with $(3\bar{2})_2$
  and $(5\bar{2})_2$ stacking sequences, respectively.}
\begin{tabular}{ c c c c c c }
\hline\hline \multicolumn{5}{c}{Lattice parameter } \\ \hline
Structure  & a (\r{A}) & b (\r{A}) & c (\r{A}) & $\gamma$ in degree & $\Delta{V}/V$(\%)  \\ \hline
10M$(3\bar{2})_2$ &  4.36  & $5\times$4.27   & 5.35  &  93.3  & 0.20\\
14M$(5\bar{2})_2$  &  4.36  &  $7\times$4.27  &  5.36 &  96.3 & 0.34 \\
14M[expt.]~\cite{SinghAPL10}      &  4.25  &  $7\times$4.13  &  5.36 &  93.1 & \\
\hline\hline
\end{tabular}

\end{table}

We next  calculated the structural parameters of 10M and 14M modulated structures by allowing full monoclinic variation
of the constructed supercells. The monoclinic structures are obtained by displacing the atomic layers in a different manner
on top of the periodic modulations. For the 10M structure, three atomic planes are shifted in one particular direction while
two atomic planes are shifted in the opposite direction, generating a $(3\bar{2})_{2}$ stacking sequence [Fig. \ref{struct}(e)]. In the case of the
14M structure, the $(5\bar{2})_{2}$ stacking sequence was chosen, as shown in Fig. \ref{struct}(h). These $(3\bar{2})_{2}$ and $(5\bar{2})_{2}$ stacking sequences are usually considered for 10M and 14M structures, respectively~\cite{ZayakPT05,GrunerPSSB14},
although Zayak {\em et al.}~\cite{ZayakPT05} showed that a different sequence can also be stabilized for Ni$_{2}$MnGa.
In the case of Mn$_{2}$NiGa, the initial shifts of the planes were chosen from the experimentally reported monoclinic angle
$\beta=93.1^{\circ}$ for a 14M  structure, although there is no result reported about the stacking of
atomic planes~\cite{SinghAPL10}.
Thus, the initial shift of the two consecutive lattice planes was taken to be 0.356\r{A} in our calculations.
The structures were relaxed with respect to the ionic positions, and the shape and volume of the supercells.
The relaxed structural parameters for these two structures are given in Table\ \ref{table2}.
Our calculated lattice parameters for the 14M structure have small differences with the experimentally
reported parameters for Mn$_2{}$NiGa. This difference  might be due to the presence of  extra 3.5$\%$ Mn
content in the experimental sample~\cite{SinghAPL10} or might be due to the stacking sequence we chose in our calculations.

The $(c/a)$ ratio in the martensitic phase is related to the maximum theoretically achievable MFIS as $\left|1-c/a\right|$.
In case of Ni-rich Ni-Mn-Ga, the experimental MFIS value was found to be almost close to the theoretical maximum MFIS for
10M (c/a$\approx$0.94) and 14M (c/a$\approx$0.90) structures.
From the results on the structural parameters, we conclude that for Mn$_2{}$NiGa, the maximum achievable MFIS for wavelike modulated structure is 11$\%$ (c/a$\approx$0.89) and that for monoclinic structure is 13$\%$ (c/a$\approx$0.87),
both higher than those in Ni-rich Ni-Mn-Ga systems.
    To date, there is only one measurement of the MFIS in
    Mn$_{2}$NiGa available for comparison with our theoretical
    prediction reporting a  $4 \%$ MFIS for the NM structure of
    Mn$_{2}$NiGa~\cite{LiuAPL05}. This should be compared with a maximum
    of $0.17 \%$ MFIS obtained in the NM tetragonal phase of Ni$_2$MnGa
    single crystals with composition ratio close to $2:1:1$~\cite{ChernenkoAPL09}.
    In Ref.\ \onlinecite{LiuAPL05}, it was observed that the reported
    $4\%$ MFIS was nonsaturating at a rather high magnetic
    field of 1.8 T. This raises the expectation that with sufficiently higher
    magnetic fields considerably larger MFIS may be obtained in Mn$_{2}$NiGa.
    In turn, experiments that observed modulated phases of
    Mn$_{2}$NiGa~\cite{SinghAPL10,BrownJPCM10} did not report MFIS.
    This is in contrast to the
    prototype Ni$_{2}$MnGa, where strain as large as nearly $10 \%$
    was obtained for the 14M orthorhombic structure with a magnetic
    field of only 0.5 T~\cite{SozinovAPL02,SozinovIEEE02}.

 The larger MFIS obtained in the modulated martensite phases of
    Ni$_{2}$MnGa are attributed to the low twinning stresses in comparison to
    the magnetic stress due to the presence of adaptive modulations and mesoscopic twin boundaries
    ~\cite{KaufmannPRL10,KaufmannNJP11}.
    Consequently, their absence explains the larger twinning stress and
    low MFIS in the NM martensite phase of Ni$_{2}$MnGa~\cite{SozinovJPIVF04,SoolshenkoJPIVF03}.
    A recent work of Sozinov {\it et
      al.}~\cite{SozinovAPL13},however, suggests that a MFIS, as
    large as $12\%$ can be obtained in the NM martensite phase if the
    twinning stress can be lowered by lowering the tetragonal
    distortion, which was realized in off-stoichiometric samples after
    simultaneous codoping of Co and Cu.
    A similar strategy might be successful for  Mn$_{2}$NiGa, as well, as
    as the observed MFIS so far, is already significantly
    higher than Ni$_{2}$MnGa.
    Detailed quantifications of twinned and
    magnetic stress 
    are, however, beyond the scope of the present work.

\subsection{Energetics of the  modulated structures}
Figure.\ \ref{en-tot} shows the  total energy curves, as a function of the tetragonality, denoted by the value of $(c/a)$,
for different modulated and the NM tetragonal structures of Mn$_{2}$NiGa considered  here.
The zero energy ($c/a=1$) in the NM curve corresponds to the high-temperature cubic structure.
In case of the NM structure, two energy minima are observed, one corresponds to the lowest energy L1$_0{}$ structure
at $c/a=1.28$ and another corresponds to the aforementioned metastable minimum at $c/a=0.93$.
The modulated structures 6M, 10M, 14M$_{3/7}$ and 14M$_{2/7}$, obtained by shuffling lattice planes along the (110)
directions in the cubic structure, have lower energies as is seen from the total energies of
different modulated structures at $c/a=1$ (Fig.\ \ref{en-tot}).
At $c/a=1$, the 6M structure has the lowest energy, the energy of the structure being lower by
about 3 meV per atom with respect to that of the cubic structure. 
It is to be noted that for Ni$_2{}$MnGa,
it was the 10M structure that had the lowest energy at $c/a=1$~\cite{HickelAEM12},
while only the 6M has a metastable energy minimum close to the pseudocubic case~\cite{GrunerPSSB14}.
After relaxation of the lattice parameters,
while keeping the orthorhombic shape of the cell, all modulations gain significantly in energy by
increasing the pseudotetragonal distortion.
Therefore, we propose that a pseudocubic 6M premartensite would be unstable against
further structural distortions. With the orthorhombic constraint for the simulation cell,
the minimum energy state is obtained for the 14M$_{2/7}$ modulation.
This is in agreement with the results of Brown {\em et al.} obtained at low temperature~\cite{BrownJPCM10}. Relaxations of the shapes of the cells convert the 10M and 14M pseudotetragonal structures to monoclinic 10M$(3\bar{2})_{2}$ and 14M$(5\bar{2})_{2}$ structures by lowering energies further (shown in Fig.\ \ref{en-tot} by filled triangles).
Once again, the 14M$(5\bar{2})_{2}$ structure has the lowest total energy, in agreement with the observation of a monoclinic 14M structure by Singh {\em et al.}~\cite{SinghAPL10}.
The hierarchy of energetics for the nanotwinned structures with monoclinic symmetry is
the same as obtained from DFT calculations for Ni$_{2}$MnGa~\cite{GrunerPSSB14}.
However, there is a noticeable difference.
For Ni$_{2}$MnGa, the energy difference between the relaxed 14M structure
(the modulated structure with the lowest energy)
and the NM tetragonal structure was nearly zero whereas in the present case of Mn$_{2}$NiGa,
the difference is about 18 meV per formula unit, significantly larger than that in the case of Ni$_{2}$MnGa.
This indicates a significant energy penalty associated with the formation of nanotwin interfaces in Mn$_{2}$NiGa
as opposed to Ni$_{2}$MnGa, where this interface energy turns out to be very small.
According to the concept of adaptive martensite of Khachaturian {\em et al.}~\cite{adaptivePRB91}, a
large interface energy can effectively
inhibit the formation of adaptive microstructures on the nanoscale.
This may, however, be overcome by externally applied stress, which conforms with
the experimental observation of Singh {\em et al.}~\cite{SinghAPL10}
that the observation of modulated structures in Mn$_{2}$NiGa are dependent on the residual stress.
Other experiments indicate the formation of secondary phases or precipitates depending on the
annealing conditions~\cite{ZhangScripta08,CaiAPL08,BrownJPCM10}.
Respective changes in local composition due to
segregation and chemical disorder may alter the energy landscape considerably. For instance,
it has been reported that at least the magnetic moments are significantly affected by
chemical disorder and structural distortion~\cite{DSouzaJoPC14,SchaeferJoPD16,PaulJAP14}.

Our present study is restricted to the case of single-phase, fully stoichiometric systems,
where we can conclude from the $T=0$ energy landscape, that due to the lack of energy barriers
aside from microstructure, the system should evolve 
downhill into a martensitic state
with significant tetragonal or orthorhombic distortion
along a comparatively steep transformation path, once a specific modulation
is established. 
Thus, we interpret the absence of a corresponding energy minimum as the signature of prevention of
the formation of a stable premartensite, in agreement with experiment.
Further, once a modulation with tetragonal or orthorhombic lattice parameters
is established, we can expect that it
will subsequently transform
into a nanotwinned monoclinic structure, according to the
associated energy gain of $6$ to $8$ meV/atom. This structure consists of
nonmodulated building blocks
separated by atomically sharp twin boundaries.
In Ni$_2$MnGa, the nonmodulated and nanotwinned arrangements are nearly
degenerate, which effectively slows down the coarsening
process~\cite{NiemannAEM12,GrunerPSSB14,Gruner2017}.
In Mn$_2$NiGa, however, the system can gain further $5$ to $7$ meV/atom
by eliminating the twin boundaries, which should speed up the kinetics of the
transition to the L1$_0$ ground state significantly, such that nanotwinned structures
might not become observable unless they are stabilized by external strain.
A concrete prediction of temperature- and stress-induced transitions
in terms of the free energy must be left open at this point, since this involves
a computationally significantly more involved estimate
of vibrational and magnetic entropy, which can play a decisive role at finite temperatures
as shown previously for Ni$_2$MnGa~\cite{UijttewaalPRL09,DuttaPRL16}.
In the next subsection, we discuss the relative stabilities of different martensitic structures,
as shown in Fig.\ \ref{en-tot} and Table \ref{table3} in terms of their electronic structures.

\begin{table}[H]
\centering
\caption{\label{table3}The equilibrium energies (in meV per atom)  of the NM tetragonal (L1$_{0}$) and different
  modulated structures relative to the energy of the high-temperature cubic phase.
  The zero energy is considered to be that of the cubic structure ($A$).
  The corresponding c/a ratios are given. For monoclinic structures, c/b ratios are given in parentheses.}
\begin{tabular}{c@{\hspace{0.5cm}}c@{\hspace{0.5cm}}c@{\hspace{0.5cm}}}
\hline\hline
& c/a$_0$ & Energy(meV/atom) \\ \hline             
E$_{A-6M}$ &  0.89  &  -12.00     \\
E$_{A-10M}$ &  0.88  &  -14.47     \\
E$_{A-14M_{3/7}}$ &  0.89  &  -14.35     \\
E$_{A-14M_{2/7}}$ &  0.88  &  -18.42     \\
E$_{A-10M(3\bar{2})_2}$ &  0.87(0.88)  &  -20.71     \\
E$_{A-14M(5\bar{2})_2}$ &  0.87(0.89)  &  -22.69    \\
E$_{A-L1_{0}}$ &  1.28  &   -27.37    \\
\hline\hline
\end{tabular}

\end{table}

\subsection{Electronic structures of the modulated phases}
\begin{figure}[t]
\centerline{\hfill
\psfig{file=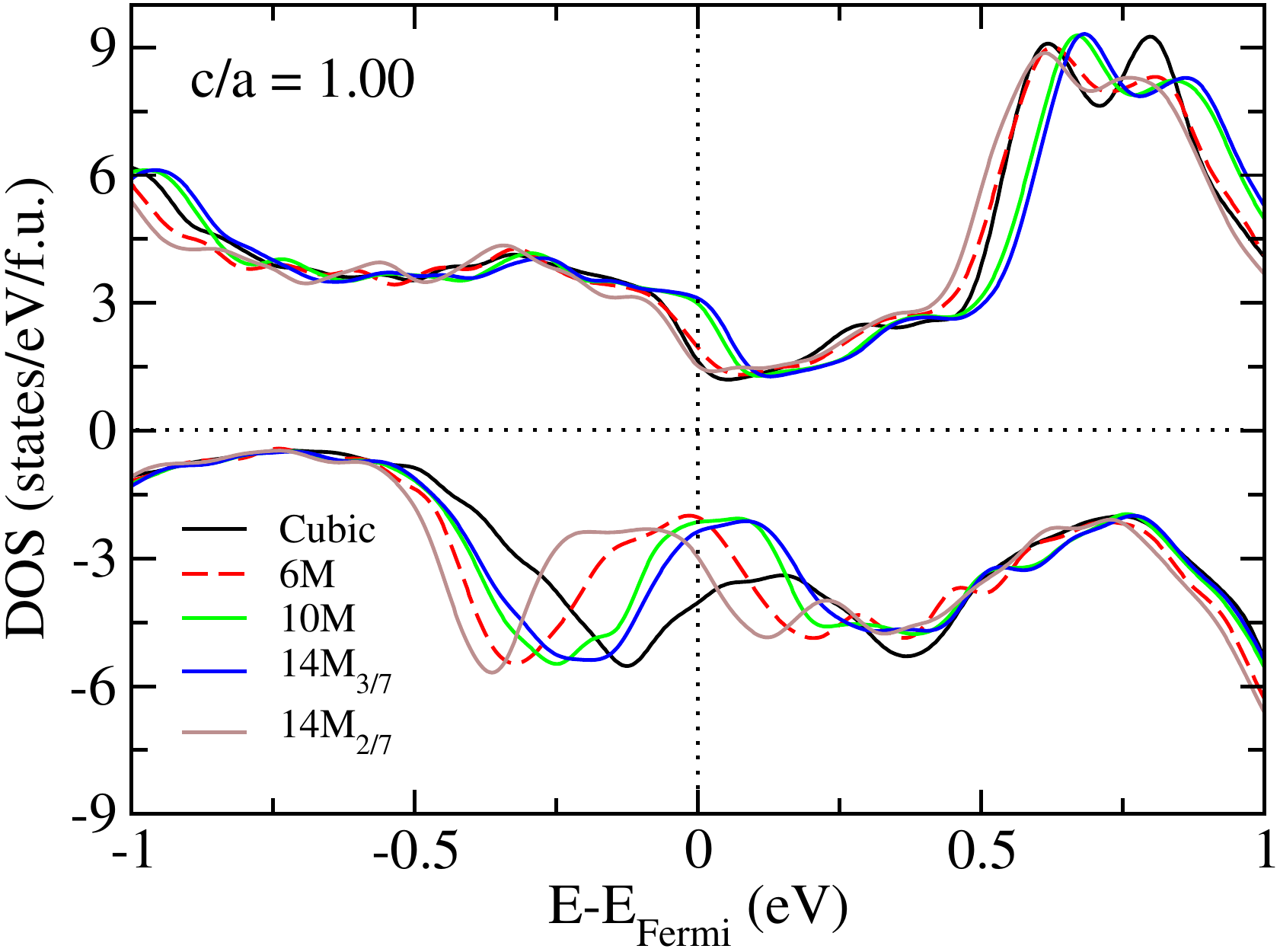,width=0.40\textwidth}\hfill}
\caption{Total density of states of different pseudocubic modulated phases. The total density of states in the cubic structure is shown for comparison.}
\label{dos-tot}
\end{figure}

\begin{figure}[t]
\centerline{\hfill
\psfig{file=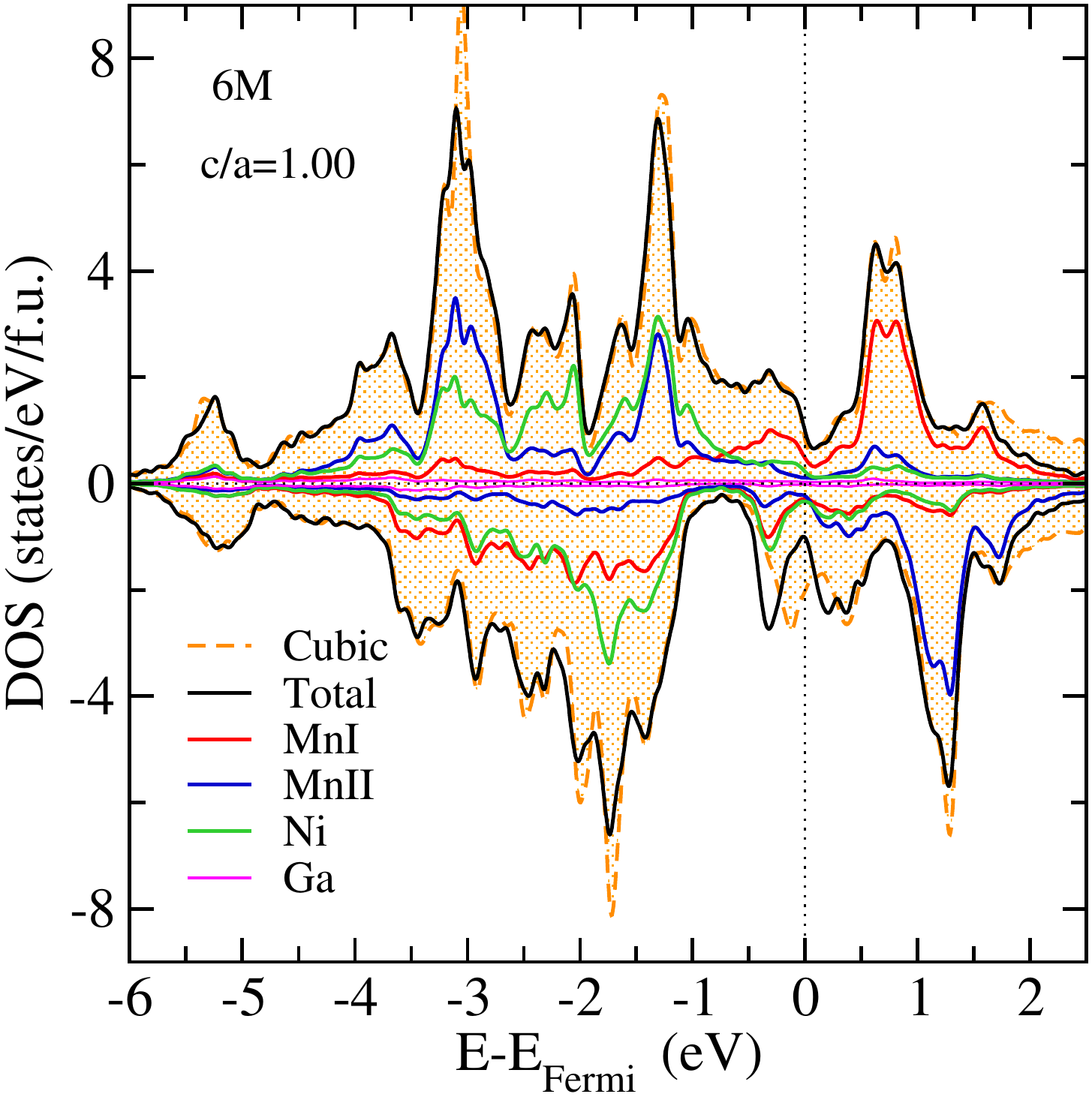,width=0.38\textwidth}\hfill}
\caption{Total and atom-projected density of states of Mn$_{2}$NiGa in the pseudocubic  6M modulated structure. The total density of states in the cubic Hg$_{2}$CuTi-type structure is shown for comparison.}
\label{dos-6m}
\end{figure}

In this section, we show the results on the densities of states (DOS) of different modulated and NM structures
and try to understand the results on relative stabilities of the different structures by analyzing them.
We first focus on the comparisons of the densities of states of the pseudocubic structures, the modulated
6M, 10M and 14M structures with $c/a=1$, with that of the cubic structure.
The total densities of states near the Fermi level for the cubic and the four pseudocubic structures are shown
in Fig.\ \ref{dos-tot}.
Since the relative stabilities of the different structures can be understood from the features near the Fermi level,
it is enough to focus on the features in the nearby region only.
The results of Fig. \ref{dos-tot} clearly show that the major changes in the densities of states are
coming from the minority spin bands. For the cubic structure, the minority densities of states at the Fermi level
were the largest.
The lowering of the symmetry due to modulation redistributes the states at the Fermi level, which stabilizes
the modulated phases.
The particular stability of the 6M pseudocubic structure is reflected in the opening of a pseudogap in the minority band at
the Fermi level.
The direct comparison between the cubic structure and the 6M pseudocubic structure presented in
Fig.\ \ref{dos-6m} shows this more prominently.
A look at the atomic contributions reveals that the features near the Fermi level,
distinguishing the electronic structures of different phases, stem from the $d$ electrons of MnI and Ni
atoms which are occupying lattice positions, which correspond to the equivalent positions of Ni
in the case of Ni$_{2}$MnGa. In both cases these states are responsible for the features in the electronic structures near the Fermi
level.

The relative stability of the 6M, 10M and 14M modulated structures with
orthorhombic lattice parameters, obtained upon relaxation of the lattice parameters of the pseudocubic structures,
can be motivated in the same way. A more elaborate discussion on this along with the densities of states for different pseudotetragonal structure is in Ref. \onlinecite{Supplement}.

\begin{figure}[t]
\centerline{\hfill
\psfig{file=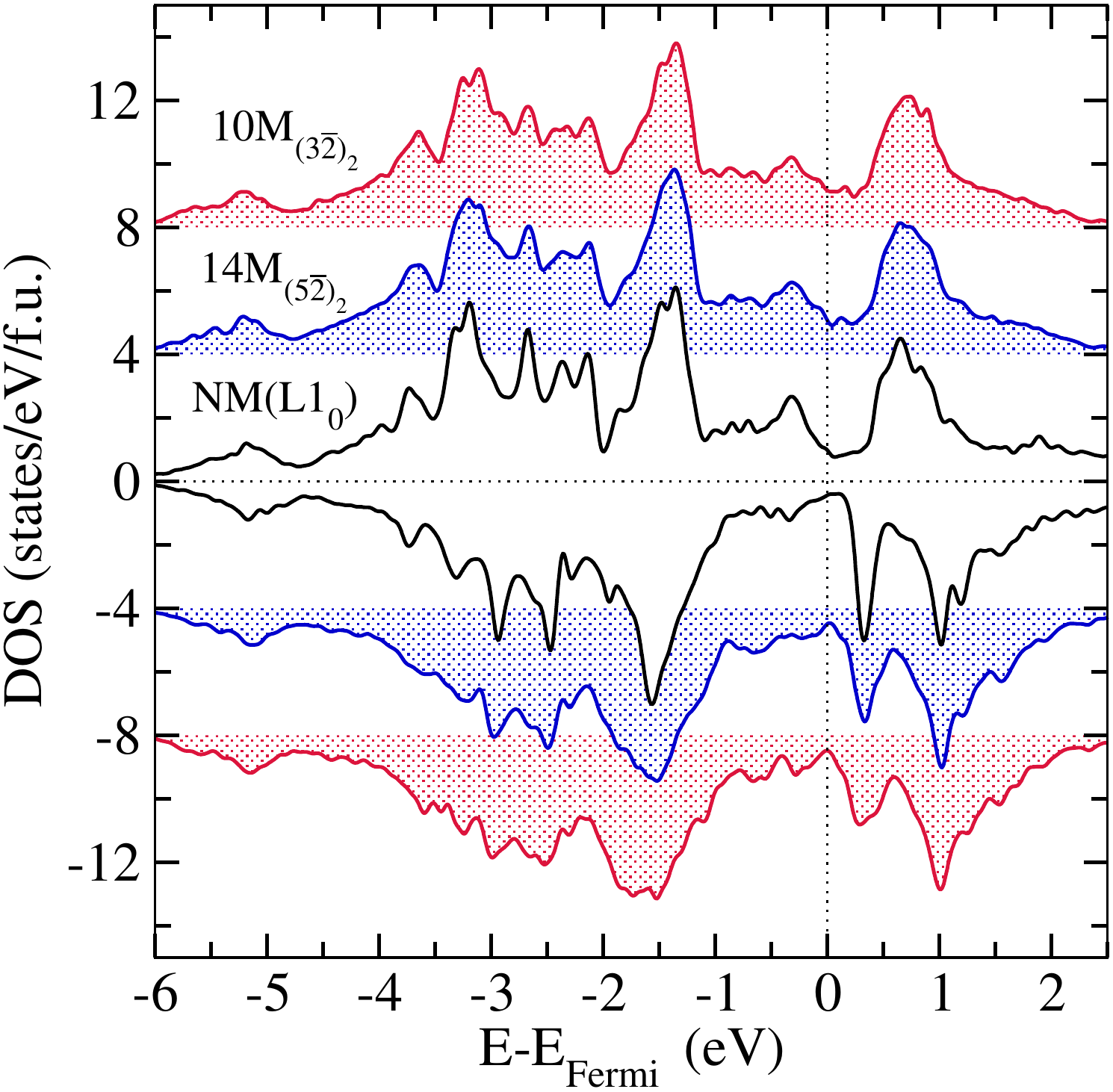,width=0.38\textwidth}\hfill}
\caption{Total density of states of  fully relaxed modulated structures. The density of states in the NM tetragonal (L1$_{0}$) structure is also plotted.}
\label{dos-mon}
\end{figure}

In Fig.\ \ref{dos-mon}, we show a comparison of the densities of states of monoclinic 10M and 14M structures along
with that of the NM tetragonal structure. Here, the overall structures of the densities of states are very similar to each other.
In particular, very few changes are observed around the Fermi level. The densities of states at the Fermi level
gradually decrease from the 10M structure to the NM structure, a trend consistent with the trends in the energetics,
thus explaining the relative stabilities of the three phases.
In Ni$_{2}$MnGa, a similar close correspondence was observed for the densities of states for the three structures.
Since the electronic density of states represents a fingerprint of the local electronic environment at each site,
this leads to the understanding that these optimized modulated structures are defected representations
of the NM martensite, the so-called nanotwinned martensites.
This, together with the extremely low values of $\gamma_{twin}$, the interface energies corresponding to the
martensitic twin boundaries, led to the conclusion that the Ni$_{2}$MnGa is a paradigmatic example of an adaptive martensite~\cite{KaufmannPRL10,KaufmannNJP11,NiemannAEM12}.
For Mn$_{2}$NiGa, although the close connections between the densities of states of the fully relaxed
modulated martensites and the NM martensite are visible, the values of $\gamma_{twin}$  are one order of magnitude larger
than those in Ni$_{2}$MnGa. 
The $\gamma_{twin}$ is defined by the energy difference between the NM and the corresponding modulated structure per unit area of the interface. The energy difference between the NM and the 10M$(3\bar{2})_{2}$ and 14M$(5\bar{2})_{2}$ modulated structures of Mn$_{2}$NiGa are 26.63 meV/f.u. and 18.72 meV/f.u., respectively. So, the calculated  $\gamma_{twin}$ values for  the 10M$(3\bar{2})_{2}$ and 14M$(5\bar{2})_{2}$ modulated structures are 2.86 meV/\AA$^{2}$  and 2.02 meV/\AA$^{2}$, respectively while those for Ni$_{2}$MnGa are only 0.25 meV/\AA$^{2}$ and 0.06 meV/\AA$^{2}$, respectively. 

The similarities in the electronic structures of the fully relaxed modulated structures and the NM tetragonal structure
along with the geometric visualization of the NM tetragonal building blocks in the supercell of the modulated structures
indicate that the modulated structures can be conceived as nanotwinned structures with a high number of interfaces between
its nonmodulated twins, but the high values of $\gamma_{twin}$ hamper the realization of
adaptive martensites in Mn$_2$NiGa.
The question of whether such structures can become thermodynamically stable without additional external stress can only be addressed through the computation of the free energy and is beyond the scope of the present work. Another immediate outcome of the similarities in the electronic structures across various phases is the similarity in their element-resolved magnetic moments listed in Ref. \onlinecite{Supplement}. The moments agree well with previously reported numbers~\cite{BarmanEPL07,PaulJAP11,WollmannPRB15,DSouzaJoPC14,CaiAPL08,BlumAPL11,AhujaPRB09,SchaeferJoPD16}.
In the next subsection, we focus on understanding  the origin of the modulated martensites in Mn$_{2}$NiGa.

\subsection{Phonon instability, Fermi Surface Nesting and Generalised Susceptibility}\label{sec:FS}
The total energy curves and the analysis of the densities of states of austenite and different modulated and NM martensites of Mn$_{2}$NiGa provide us with the possible sequence of phase transformation, relative stabilities of different martensites and changes in the electronic structures as the system makes a transition from the cubic austenite to the NM martensite phase via different modulated structures. The question now remains as to why different modulated structures are observed to occur.

Phonon dispersion relations in Ni$_{2}$MnGa provided an important clue to the origin of particular modulated structures observed experimentally. An anomalous softening of the transverse acoustic (TA$_{2}$) phonon branch in [$\xi \xi 0$] direction in the cubic L2$_{1}$ phase, with the pronounced softening at $\xi\approx 0.33$, $\xi$ being the reduced reciprocal lattice vector, was interpreted to be the precursor to the martensitic transformation. DFT calculations of phonon dispersion relations for the L2$_{1}$ austenite phase in Ni$_{2}$MnGa provided completely unstable transverse acoustic mode (TA$_{2}$) along the [$\xi \xi 0$] direction with the $\xi$ value at which the instability is pronounced being close to 0.33~\cite{BungaroPRB03}. The wave vector at which the instability is maximum could be fitted into a 6M modulated structure, and thus the 6M phase was interpreted to be the premartensitic phase arising out of the phonon instability. Later DFT calculations found the phonon instability rather at around $\xi\approx 0.4$~\cite{ZayakPRB03}, which could be connected to a 10M modulated structure observed in diffraction experiments. Taking the cue from Ni$_{2}$MnGa, we have computed the phonon dispersion relations for Mn$_{2}$NiGa.

\begin{figure}[t]
\centerline{\hfill
\psfig{file=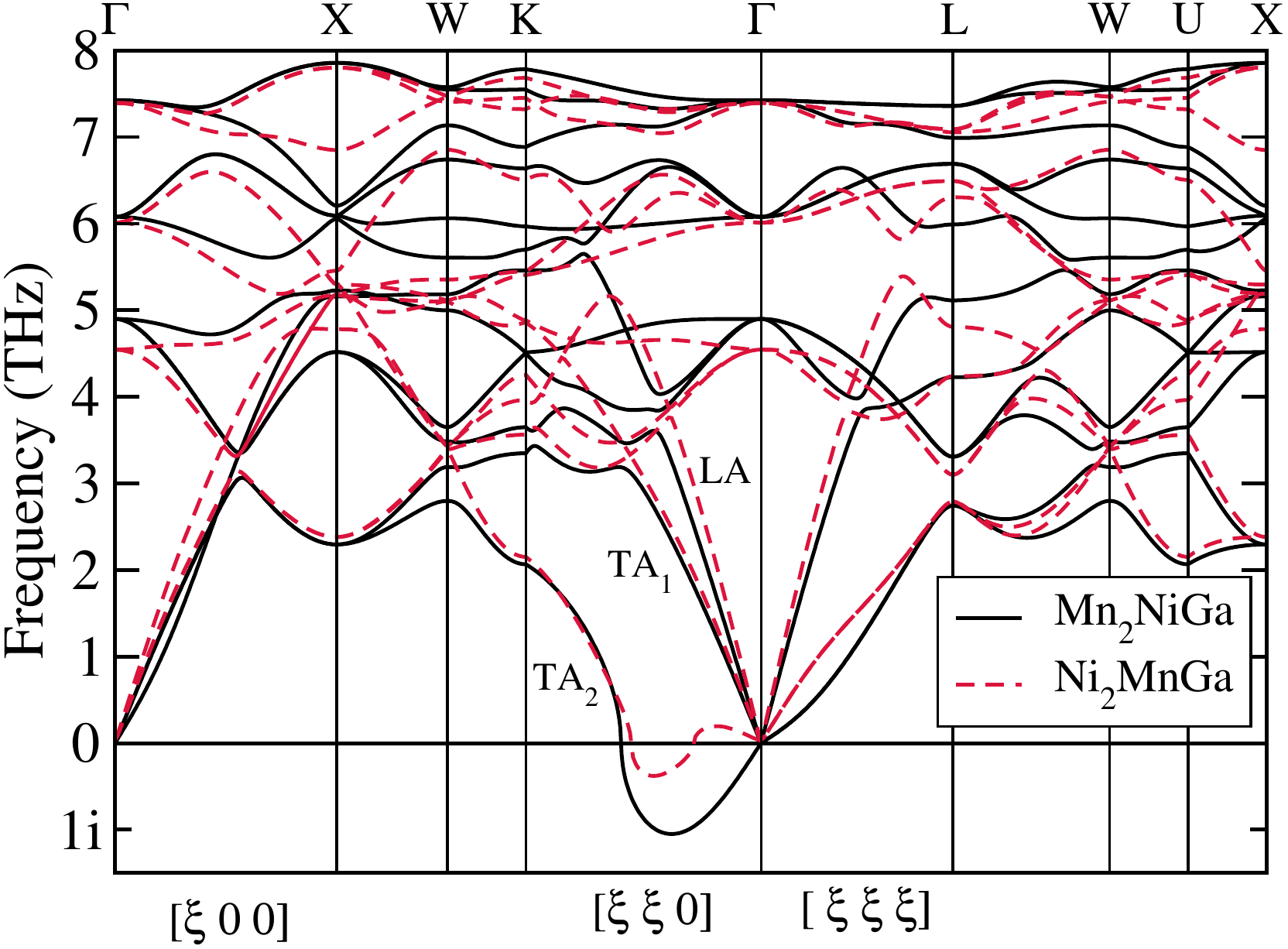,width=0.42\textwidth}\hfill}
\caption{Full phonon dispersion curve for Mn$_{2}$NiGa (solid line) and Ni$_{2}$MnGa (dashed line) in the cubic structure.
The calculated dispersion for Ni$_{2}$MnGa was taken from Ref.\ \onlinecite{EnerPRB12}.}
\label{phonon}
\end{figure}

In Fig. \ref{phonon} we plot the phonon dispersion relations for Mn$_{2}$NiGa along different symmetry directions in the entire Brillouin zone. We compare this directly with the calculated dispersion of Ni$_{2}$MnGa from Ref.\ \onlinecite{EnerPRB12}, which was obtained in a similar technical setup and agrees excellently with the dispersion measured by inelastic neutron scattering. In Mn$_{2}$NiGa, along the [$\xi \xi 0$] direction, we find the TA$_{2}$ mode to be unstable with the frequency being imaginary in the range between $\Gamma$ and $\xi=0.45$ with a pronounced instability (the minimum value of the frequency) at $\xi\approx 0.33$. The results agree very well with an earlier calculation with density functional perturbation theory (DFPT) using ultrasoft pseudopotentials~\cite{PaulJPCM15}, except a small deviation regarding the range for which the frequency of the TA$_{2}$ mode is imaginary. This small deviation might be due to use of different pseudopotentials and different parameters in DFPT~\cite{PaulJPCM15}.
We do not find any instability in any of the other symmetry directions. The comparison with Ni$_{2}$MnGa shows that the dispersion relations in the two materials are very similar, highlighted by the single instability along the [$\xi \xi 0$] direction, and differing only in the $\xi$ value at which the instability is most pronounced. This implies that the phonon related physics would be very similar for the two systems. One immediate consequence of the phonon instability in Mn$_{2}$NiGa would be to associate it with the 6M modulated structure which is found to be the most stable pseudocubic structure in our calculations, much the way it was done in case of Ni$_{2}$MnGa~\cite{BungaroPRB03,ZayakPRB03}.
However, deeper understanding into the electronic origin of the modulated structures needs to be gained from further analysis of the phonon dispersions along the given direction.

\begin{figure}[t]
\centerline{\hfill
\psfig{file=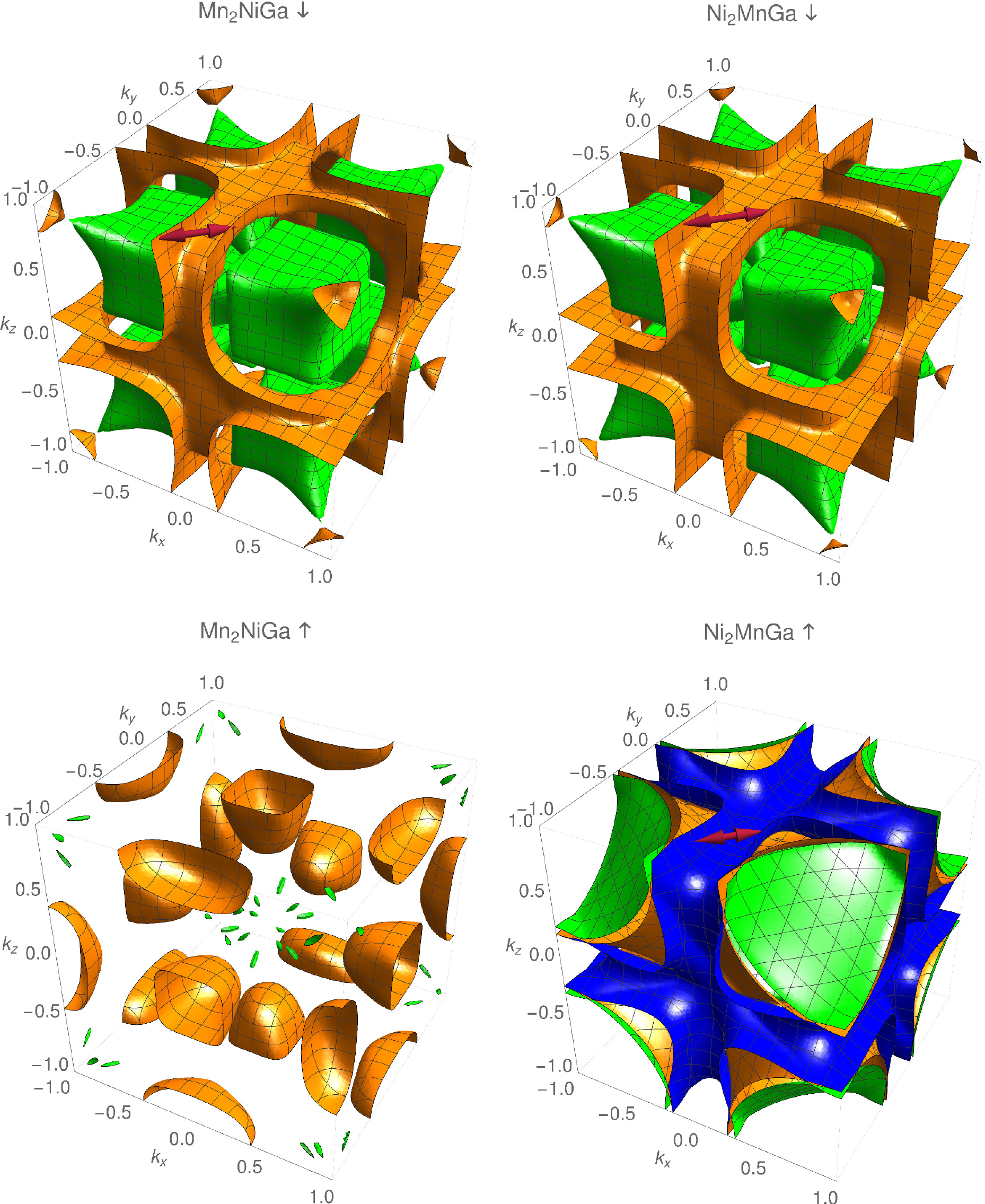,width=0.45\textwidth}\hfill}
\caption{Minority spin (top row) and majority spin (bottom row) Fermi surfaces
  of Mn$_2$NiGa (left column) and Ni$_2$MnGa (right column) in
  the extended zone scheme, represented by a cubic box with the edge length
  $2\pi/a$. The center and the corners of the box refer to the $\Gamma$ point.
Each band crossing the Fermi energy (three in the case of
the majority spin channel and two for the minority spin channel) is indicated
by an individual color (green: 17$^{\rm th}$ spin-up and the 13$^{\rm th}$ spin-down band of
Ni$_2$MnGa and 14$^{\rm th}$ spin-up and 13$^{\rm th}$ spin-down band of Mn$_2$NiGa;
orange: 18$^{\rm th}$ spin-up and 14$^{\rm th}$ spin-down band of
Ni$_2$MnGa and 15$^{\rm th}$ spin-up and 14$^{\rm th}$ spin-down band of Mn$_2$NiGa;
blue: 19$^{\rm th}$ spin-up band of Ni$_2$MnGa). 
The Fermi surfaces agree excellently with previous calculations for
Mn$_2$NiGa\protect~\cite{BarmanEPL07,PaulJPCM15} and
Ni$_2$MnGa.\protect~\cite{BungaroPRB03,OpeilPRL08,EntelMSF08,Entel08MRS,HaynesNJP12,Siewert12AEM,Entel12Book}
The red arrows indicate possible nesting vectors according to the respective
  first maxima of $\chi^{\prime}_0(\vvec{q},0)$ at $\vvec{q}$$\,=\,$$\xi_1\,[1\,1\,0]$ as
  calculated in this work.}
\label{FS}
\end{figure}

%
%

\begin{figure*}[t]
\centerline{\hfill
\psfig{file=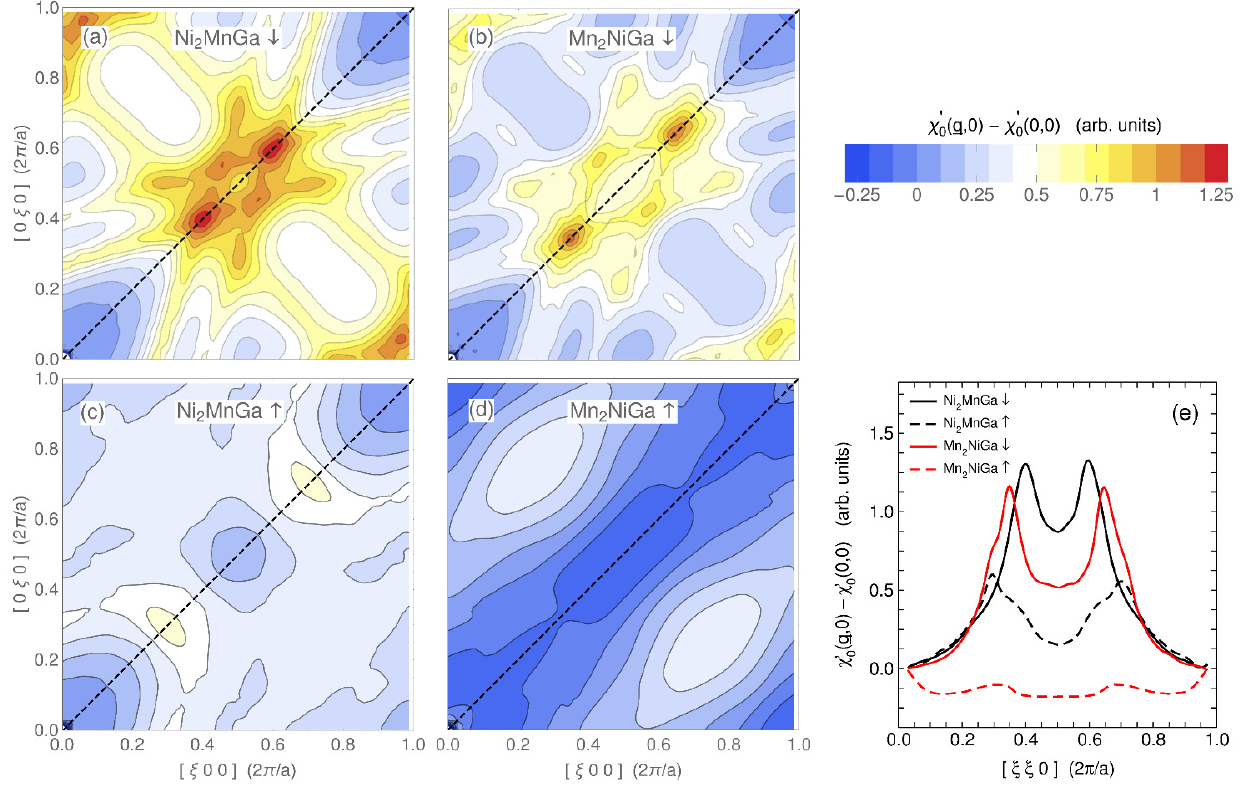,width=0.90\textwidth}\hfill}
\caption{  Real part of the generalized susceptibility in the low-frequency limit
  $\chi^{\prime}_0(\vvec{q},0)$
  of Ni$_2$MnGa and Mn$_2$NiGa for wave vectors $\vvec{q}$
in the $[1\,0\,0]$-$[0\,1\,0]$ plane: (a) and (b) show the minority spin
contribution, (c) and (d) the majority spin channel.
Panel (e) shows the cross sections along the $[1\,1\,0]$ direction
indicated by the dashed lines in (a)$-$(d).
In all cases, a moderate electronic temperature of $T=50$ K was assumed to smoothen the features.
The long-wavelength limit $\chi^{\prime}_0(0,0)$ has been subtracted, such that the absolute
magnitude of the features represented by the color coding can be compared.
The difference between two neighboring contour lines is 0.1 (arbitrary units)
in (a)-(d).}
\label{fig:Chi}
\end{figure*}

The softening of acoustical modes in martensitic materials at a finite fraction $\xi$ of
a reciprocal lattice vector $\vvec{q}$
in a specific direction
is frequently discussed in terms of electron-phonon coupling.
The contribution from electron-phonon coupling to the dynamical matrix
is given by the
expression~\cite{VarmaPRL77,VarmaPRB79}
\begin{eqnarray}
D_{\rm el-ph}^{\alpha\beta}(\vvec{q}) & = & \sum_{\vvec{k},m,n}\,
\frac{f(\varepsilon_{\vvec{k},m}) -
f(\varepsilon_{\vvec{k}+\vvec{q},n})}{
\varepsilon_{\vvec{k}+\vvec{q},n} -\varepsilon_{\vvec{k},m}} \quad\times\nonumber \\
& & \times\quad
g^{\alpha\ast}_{\vvec{k},m,\vvec{k}+\vvec{q},n}\,g^{\beta}_{\vvec{k}+\vvec{q},n,\vvec{k},m}
\,,
\label{eq:ElPh}
\end{eqnarray}
where $f$ is the Fermi distribution function, $\varepsilon_{\vvec{k},m}$ are
the energies of band $m$ corresponding to the reciprocal lattice vector $\vvec{k}$, and
$\alpha$, $\beta$ refer to the Cartesian directions. The $g$ denote
the matrix elements describing the strength of the electron-phonon coupling.
As these are difficult to calculate and in many cases only little variation is
expected, they are often treated as constant.
In the constant matrix element approximation,
one can obtain, from the phonon self energy,
the so-called generalized susceptibility
$\chi^{\prime}_0(\vvec{q},0)$~\cite{PickettPRB77}, which is the real part of the bare electronic
susceptibility
\begin{equation}
\chi_0(\vvec{q},\omega) = \sum_{\vvec{k},m,n}\,
\frac{f(\varepsilon_{\vvec{k},m}) -
f(\varepsilon_{\vvec{k}+\vvec{q},n})}{
\varepsilon_{\vvec{k}+\vvec{q},n} -\varepsilon_{\vvec{k},m}-\hbar \omega
- i \delta}
\label{eq:ChiOmega}
\end{equation}
in the low phonon frequency limit $\omega\to 0$.
The denominator of $\chi^{\prime}_0(\vvec{q},0)$ is the difference in the occupation
of initial and final states connected by a wave vector $\vvec{q}$, which becomes maximum when the former
are occupied and the latter empty. Therefore peaklike features
in the generalized susceptibility can be regarded as indication for an
electronic instability which is related to the topology of the Fermi surface.
From the imaginary part $\chi^{\prime\prime}(\vvec{q},\omega)$, one can derive the quantity
\begin{equation}
\lim_{\omega\to 0}\,\frac{\chi^{\prime\prime}_0(\vvec{q},\omega)}{\omega}
=  \sum_{\vvec{k}}\,\delta\left(\varepsilon_{\vvec{k}}-\varepsilon_{\rm
    F}\right)\,\delta\left(\varepsilon_{\vvec{k}+\vvec{q}}-\varepsilon_{\rm
    F}\right)\,.
\label{eq:Nestdens}
\end{equation}
It is termed the ``nesting density'' since it provides a measure for
nesting regions of the Fermi surface, which are connected by a rigid vector $\vvec{q}$.
The nesting density is straightforwardly obtained from the Fermi surface alone,
whereas
$\chi^{\prime}_0(\vvec{q},0)$ also contains contributions from bands farther
away from the Fermi surface. These are not effectively suppressed
since their contributions scale inversely to
the energy difference between occupied and unoccupied states. Indeed, this may be significant in the sense that their extremal features do not
necessarily coincide.
One example is the so-called hidden nesting observed in
purple bronze, which can occur, although obvious nesting features are absent according
to avoided crossing of bands right at the Fermi surface~\cite{WhangboSC91}.
Consequently, the use of Eq.\ (\ref{eq:Nestdens}) as a replacement for
$\chi^{\prime}_0(\vvec{q},0)$ has been criticized~\cite{JohannesPRB08}.

The connection between Fermi surface nesting and the transition to a
premartensitic modulated structure appears to be well established for the case
of the conventional shape memory alloys (see Ref.\ \onlinecite{KatsnelsonPT94} for a
discussion).
For the magnetic shape memory system  Ni$_2$MnGa,
a first systematic discussion of band structure, Fermi surface, and generalized
susceptibility was provided by Velikokhatnyi and Naumov
~\cite{VelikokhatnyiFTT99}.
The authors found that in the minority spin
$\chi^{\prime}(\vvec{q})$, there were two  pronounced maxima along [110] (in units of $\pi/a$, where
$a$ is the lattice constant), one at $\xi_1=0.43$ and a second at $\xi_2=0.59$.
Due to the discrepancy with the experimentally observed phonon softening
at experimental value $\xi\approx 0.33$, the authors suggested that the transition to the 6M premartensite
should rather be of anharmonic origin following
the suggestion of Gooding and Krumhansl~\cite{GoodingPRB88,GoodingPRB89}, while
$\xi_1$ corresponds to the incommensurate 10M martensite observed in diffraction experiments.
Lee, Rhee, and Harmon~\cite{LeePRB02} pointed out that a peak corresponding to the 6M premartensite
exists in the $\chi^{\prime}(\vvec{q},0)$ of the 17$^{\rm th}$ majority spin band.
The position of this feature is fine tuned through the reduction of the Ni exchange splitting and  eventually
matches the observed 6M modulation due to the increasing magnetic disorder at finite temperatures, which was confirmed later on by similar calculations of Haynes and co-workers~\cite{HaynesNJP12}. Bungaro, Rabe, and Dal Corso~\cite{BungaroPRB03} directly compared the Fermi surface with the full phonon dispersion of austenite Ni$_2$MnGa calculated
by means of density functional perturbation theory~\cite{DalCorsoPRB97}.
Their calculations reveal an instability in the [110] direction which corresponds to the
nesting vectors obtained from cross sections of the Fermi surface.
Later work~\cite{ZayakJPCM03,ZayakPRB05,OpeilPRL08,EntelMSF08,HaynesNJP12}
confirmed this picture, which coincides well with the nesting
present in the minority spin Fermi surface shown in Fig. \ref{FS}.
The shape of the minority spin Fermi surface of Ni$_{2}$MnGa
has been verified experimentally using the positron annihilation
technique~\cite{HaynesNJP12}.
An analogous relation between Fermi surface nesting and phonon instability has also been
proposed for Mn$_2$NiGa~\cite{BarmanEPL07,PaulJPCM15}, while calculations of
$\chi^{\prime}_0(\vvec{q})$ are missing for this system, so far.
The antiparallel orientation of the spin moments of Ni and MnI allows for
a perfect hybridization of their minority spin states.
In consequence the minority Fermi surfaces of Ni$_2$MnGa and Mn$_2$NiGa are nearly identical.
The only visible difference is the
smaller distance of the (orange) plane sheets forming the
narrow rectangular channels in Mn$_2$NiGa, which originates from
the e$_u$-peak being slightly closer to the Fermi level.
In turn, MnI does not occupy all states in the majority channel, which reduces hybridization and the majority spin Fermi surfaces of Ni$_2$MnGa and Mn$_2$NiGa
do not resemble each other at all, as can be seen from the lower panels of
Fig.\ \ref{FS}. From this, we can conclude that mechanisms derived from the electronic structure of Ni$_2$MnGa should be transferable to Mn$_2$NiGa, if they involve the minority
spin states only. In contrast, effects connected to features of the majority spin electronic
structure in Ni$_2$MnGa should not be observed in Mn$_2$NiGa.

In Fig.\ \ref{fig:Chi} we present a detailed comparison of $\chi^{\prime}_0(\vvec{q},0)$
between both systems,  Ni$_2$MnGa and  Mn$_2$NiGa, for both spin channels.
Unlike previous studies on Ni$_2$MnGa, we take into account
vectors $\vvec{q}$ in the whole $[1\,0\,0]$-$[0\,1\,0]$ plane. We consider all bands present
in an interval of $\pm\,0.5\,$eV around the Fermi level, which are in particular
the 17$^{\rm th}$-19$^{\rm th}$ spin-up and the 12$^{\rm th}$-15$^{\rm th}$ spin-down bands for
Ni$_2$MnGa and the
12$^{\rm th}$-15$^{\rm th}$ spin-up and the 14$^{\rm th}$-17$^{\rm th}$ spin-down
bands for Mn$_2$NiGa (not counting the Mn- and Ni-$3p$ and Ga-$3d$ semicore states).
To allow for direct comparison, the scale of all panels in Fig.\ \ref{fig:Chi}
is the same. This was realized by subtracting  $\chi^{\prime}_0(0,0)$,
following the argument that
for $\vvec{q}\to 0$, the electronic and ionic contributions of the dynamical matrix
cancel each other~\cite{EschrigPSSB73}.
The cross section along [110] shown in Fig.\ \ref{fig:Chi} (e) shows pronounced
maxima at $\xi_1=0.4$ and  $\xi_2=0.6$ in the minority spin channel and
at $\xi_1=0.3$ and  $\xi_2=0.7$ in the majority channel for Ni$_2$MnGa, which reproduces the
trend previously reported by Lee {\em et al.}~\cite{LeePRB02},
in particular the much smaller magnitude of the features in the majority compared
to the minority channel.

The small differences in the numerical values of
Refs.\ \onlinecite{VelikokhatnyiFTT99} and \onlinecite{LeePRB02} may be related to the
use of the GGA in the present study as opposed to the LDA exchange correlation used previously.
For Mn$_2$NiGa, the minority channel shows, as expected, a
very similar structure, with slightly shifted and lesser pronounced
maxima at $\xi_1=0.35$ and  $\xi_2=0.65$.
As depicted in Fig.\ \ref{fig:Chi} (a,b), these maxima are indeed confined to $\vvec{q}$ along $[1\,1\,0]$.
If we identify the maximum at smaller $\xi$ with the soft TA$_2$ phonon, this
fits very well to the instability being observed at smaller wavelength in Mn$_2$NiGa.
The close similarity between both systems
in the minority spin $\chi^{\prime}_0(\vvec{q})$ extends to the entire $[1\,0\,0]$-$[0\,1\,0]$ plane and includes also the maxima
close to $[1\,0\,0]$ and the satellites around $[\frac{1}{2}\,\frac{1}{4}\,0]$.
%
%
In contrast, the generalized susceptibility shows no resemblance between both system in the majority spin channel.
In Mn$_2$NiGa, the majority spin  $\chi^{\prime}_0(\vvec{q},0)$ is essentially featureless, while in
Ni$_2$MnGa, we find the above-mentioned maximum in the $[1\,1\,0]$ direction around $\xi_1=0.3$,
which Lee {\em et al.} related to the premartensite modulation~\cite{LeePRB02}.
On the premise that the modulation in martensite and premartensite
could be of different origin this alone is not
necessarily a contradiction, since in experiment a premartensitic
modulation has not been reported for Mn$_2$NiGa so far.

The red arrows in Fig.\ \ref{FS} show exemplarily the vectors belonging
to the maxima at $\vvec{q}$$\,=\,$$\xi_1\,[1\,1\,0]$.
As discussed above, maxima of the generalized susceptibility
do not necessarily need to indicate the most
frequent nesting vectors characterized by $\chi^{\prime\prime}_0(\vvec{q},0)$.
Nevertheless, they appear to achieve this quite well for the minority channel
of both Ni$_2$MnGa and Mn$_2$NiGa.
Here, nesting is expected between the walls of the rectangular
flat channel structure formed by the 14$^{\rm th}$ band
(indicated by the arrows in the top row of Fig.\ \ref{FS}), 
as well as between the parallel surfaces of the cube like structures formed from the 13$^{\rm th}$ band (which are
hidden beneath the sheets of the 13$^{\rm th}$
bands)~\cite{VelikokhatnyiFTT99,BungaroPRB03,EntelMSF08,HaynesNJP12,BarmanEPL07,PaulJPCM15}.
In the majority channel of Ni$_2$MnGa, nesting was reported to occur
mainly within the 17$^{\rm th}$ band
between the flattened surfaces of the central hole pocket
around $\Gamma$ and its periodic images
(again hidden in Fig.\ \ref{FS} by the surfaces 18th and 19th band)
~\cite{LeePRB02,HaynesNJP12},
while some nesting is also present in the other bands.

Conversely, the importance of Fermi surface nesting for the systems
might be assessed by comparing
the electronic density of states
between the dynamically unstable cubic $L2_1$ phase and the
energy-minimizing modulation in pseudocubic structure in a cell with 
$L2_1$ lattice parameters.
Since the redistribution of states at the Fermi level is thought
to drive the transition to the martensitic modulation in the spirit of a Peierls instability,
a condensed soft phonon should remove the nesting parts of the Fermi surface,
accompanied by a characteristically
reduced DOS around the Fermi level.
This is indeed the case for the minority spin channel
of Mn$_2$NiGa as shown in Fig. 5 (see Ref. \onlinecite{Supplement}
for other modulations) in a very similar fashion to that for the minority channel of Ni$_2$MnGa
(Ref. \onlinecite{Habil12} and Ref. \onlinecite{Supplement}).
In turn, the majority channel DOS remains essentially unaltered at the
Fermi level, even for Ni$_2$MnGa, which underlines the importance of the
minority spin $e_u$ states for the electron-phonon coupling in both cases.
Thus the
comparison of both systems yields a strong support for the view
that the softening of the $[1\,1\,0]$ TA$_2$ acoustic
branch observed in the ground-state phonon calculations is essentially
nourished by the nesting features of the
minority spin Fermi surface. A final discussion of this issue
requires, however, the first-principles evaluation of the respective
electron-phonon coupling matrix elements, which is beyond the scope of the present work.

\section{CONCLUSIONS}

Using density functional theory based calculations, we have presented a comprehensive study of the structural
and electronic properties of the recently found, comparatively less explored shape memory alloy Mn$_{2}$NiGa.
Our calculations confirmed the existence of the experimentally observed 14M modulated martensite phases with
two different symmetries. Our calculations revealed the possible existences of other modulated martensites,
6M and 10M with orthorhombic symmetry and 10M(${3\bar2}$) with monoclinic symmetry.
The total energy calculations indicated that the high-temperature cubic phase will be first driven towards a
intermediate 6M pseudocubic premartensite state, which should immediately decay downhill into a modulated structure with
a significant pseudotetragonal distortion. Afterwards the relaxations of the internal coordinates and the structural
parameters will stabilize other modulated structures.
This corresponds well to the fact that the complementary phonon calculations in the cubic phase
show that the cubic structure is unstable and that the instability can be mapped onto a 6M modulated structure.
Our results attribute the origin of the instability to the Fermi surface nesting features in the minority spin bands,
while no supporting contribution is encountered from the majority spin electronic susceptibility.
The striking similarities in the features of the minority band Fermi surfaces and the susceptibilities
for Mn$_{2}$NiGa and Ni$_{2}$MnGa lead us to believe that the electronic mechanism driving the martensite
transformations in these two systems are similar. The calculations of the structural parameters in
various modulated phases and the relative energies (of modulated phases) with respect to the non-modulated
ground state reveal important differences with Ni$_{2}$MnGa, the prototype magnetic shape memory alloy.
On the other hand, this also indicates that the maximum achievable MFIS in Mn$_{2}$NiGa would be greater
than that in Ni$_{2}$MnGa. 

In summary,
comparing our DFT results for Mn$_2$NiGa side by side with Ni$_2$MnGa,
we would like to sketch the following potential scenario for the 
martensitic transformation process:
In both systems, wavelike modulations in the cubic phase provide an 
energetic advantage
of a few meV/atom over the L2$_1$ austenite. This arises from the 
nesting features of the Fermi surface connected
to an enhanced electronic susceptibility at specific wave vectors along 
[110].
When temperature becomes sufficiently low, this may end in a static 
modulation of the ionic positions.
In Ni$_2$MnGa this corresponds to the premartensitic 6M state, which is 
stabilized
by vibrational entropy in a certain range of temperatures above the 
martensitic transformation~\cite{UijttewaalPRL09}.
Although the energy gain from a 6M modulation in the
pseudocubic lattice is significantly larger in Mn$_2$NiGa, this does 
not imply a stable premartensite phase, here.
The reason is that the pseudo-tetragonal wave-like 6M in Ni$_2$MnGa has a metastable energy 
minimum at $c/a=0.985$, close to 1,
whereas in Mn$_2$NiGa this condition is not fulfilled as the energy 
landscape implies a downhill relaxation
towards a state with a much larger tetragonal distortion $c/a$, which is 
similar for all modulated structures. Our calculations only consider commensurate modulations. Thus, we cannot exclude incommensurate modulations being lower in energy. But we expect such modifications to be small, since the extremal features in the generalized susceptibility and the imaginary phonon branches are rather close to the commensurate wave vectors. In addition, there is no reason to stop the relaxation process right here,
since the other modulations in the pseudotetragonal cell and even more
the monoclinic nanotwinned states are significantly lower in energy. 
According to the concept of
adaptive martensite~\cite{adaptivePRB91}, a specific twinning sequence 
will be favored since it additionally benefits from
minimizing the elastic energy at the interface to austenite, which 
dominates when the growing martensite plates are
still sufficiently small.
For Ni$_2$MnGa, recent DFT results imply that
pseudotetragonal
10M  modulation may finally
be stabilized by vibrational entropy~\cite{DuttaPRL16}, while the 
adaptive nanotwinned 14M structure
is nearly degenerate with the non-modulated L1$_0$ phase at 
$T=0$~\cite{NiemannAEM12,Gruner2017}.
Therefore, macroscopic regions of 14M can remain metastable for a very 
long time.
Eventually, the nanotwin boundaries may be eliminated in an irreversible 
twin coarsening
process through the (stress-induced) motion of twinning 
dislocations~\cite{KaufmannNJP11,PondAM12}.
We expect the significantly larger twin boundary energy in Mn$_2$NiGa to 
speed up this process, such that
the nanotwinned intermediate state might not be observed unless the 
sample is strained.
This, once again, corresponds well to the experimental observation of 
stress-induced modulated structures
in Mn$_2$NiGa~\cite{SinghAPL10}.

\section*{ACKNOWLEDGMENT}
Computational and financial support
from C-DAC, Pune, India, and the Department of Physics, IIT Guwahati, India through the DST-FIST program and
Deutsche Forschungsgemeinschaft within SPP 1599 (GR3498/3), are gratefully acknowledged.


\end{document}